# Fast Characterization of Moving Samples with Nano-Textured Surfaces

*Morten Hannibal Madsen[1,*], Poul-Erik Hansen[1], Maksim Zalkovskij[2], Mirza Karamehmedović[3], and Jørgen Garnæs[1]*

[1]Danish Fundamental Metrology A/S, Matematiktorvet 307, 2800 Kgs. Lyngby, Denmark

[2]NIL Technology ApS, Diplomvej 381, 2800 Kgs. Lyngby, Denmark

[3]Dept. of Applied Mathematics and Computer Science & Dept. of Physics, Technical University of Denmark, Matematiktorvet 303B, DK-2800 Kgs. Lyngby, Denmark

ABSTRACT We characterize nano-textured surfaces by optical diffraction techniques using an adapted commercial light microscope with two detectors, a CCD camera and a spectrometer. The acquisition and analyzing time for the topological parameters height, width, and sidewall angle is only a few milliseconds of a grating. We demonstrate that the microscope has a resolution in the nanometer range, also in an environment with many vibrations, such as a machine floor. Furthermore, we demonstrate an easy method to find the area of interest with the integrated CCD camera.

**Introduction**

An increased number of products utilizing micro/nano-textured surfaces are moving towards the commercial market. However, most conventional characterization techniques are inapplicable in a large-scale industrial environment. Imaging techniques such as atomic force microscopy (AFM), scanning electron microscopy (SEM), and confocal imaging are all very sensitive to vibrations. Isolation and damping of vibrations is a cumbersome task, if possible at all [1]. Thus, there is a high demand for new in-situ imaging techniques, especially in a production environment. Furthermore, the abovementioned characterization techniques are all highly time-consuming, which restricts their applicability in a large-scale industrial process flow.

In this paper we introduce an adapted optical microscope capable of measuring with resolution in the nanometer range on a moving sample. The system is based on the principles of optical diffraction microscopy (ODM) where the spectrum of the reflected light is studied [2, 3]. The scattering intensities are independent of the sample movement, as long as one observes an area with uniform structures. We show that one can shake the microscope or move the sample during acquisition without affecting the results of the measurements. This allows our adapted optical microscope to be integrated in a production line, to perform e.g. quality control in the nanoscale range.

Several types of scatterometers exist, including ODM [2, 4, 3], angular scatterometry [5, 6, 7], Fourier lens system [8], coherent Fourier scatterometry [9, 10, 11], white light interference Fourier scatterometry [12], and naked-eye observations [13]. Two general challenges for scatterometry are imaging of small areas and finding a specific area of interest. In this paper, we demonstrate a method to overcome both of these challenges by building the scatterometer into a conventional optical microscope. Furthermore, the spot size, which defines the imaged area, can easily be controlled by change of objective. Typical spot sizes are in the range from less than hundred microns to several millimeters.

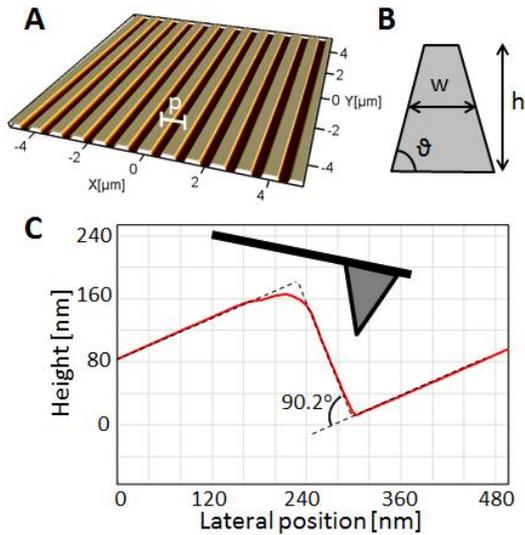

Figure 1. 1D grating in Si(100). (A) Topographic AFM image of a 1D grating with a pitch, p, of 800 nm. All axes have the same length scale. (B) Sketch of a single structure seen from the side. The definition of the height, h, width, w, and sidewall angle, $\vartheta$, is indicated in the figure. The filling factor is defined as the amount of material present compared to a uniform film with the same thickness as the height of the nanostructures. (C) Profile obtained with a tilted sample 23° in the AFM of the grating with a 1400 nm pitch. The shape and angle of the scanning AFM tip is indicated in the figure.

The scatterometer technique is suitable for structures that can be modelled using periodic or non-periodic boundary conditions. Periodic structures include one-dimensional (1D) gratings and two-dimensional (2D) arrays of structures. Both types of structures with dimensions in the micro/nano-range are entering the consumer market, and hence the need for fast and reliable characterization methods. Gratings with 1D nano-structures are used, e.g., for structural colors in the design of surfaces with iridescence [14, 15]. An AFM image of a 1D grating and typical topography parameters of interest are shown in Fig. 1. Nanowire-based solar cells [16] are typically fabricated in a well-ordered 2D array [17]. As these devices are also approaching the market, methods for fast large-area characterization with nanometer resolution are needed. The optical response for different crystal orientations [18] and the angle-dependent absorption [19] for nanowires have recently been measured, paving the road for the topological measurements. Non-periodic boundary conditions can for instance be used to model photonic crystal waveguides, Bragg mirrors, and grating couplers [20], and single structures such as micro-fluidic channels and sub-micron wires [21].

**Experimental setup**

The system, sketched in Fig. 2, is based on a Navitar optical microscope (12x zoom) equipped with a 5W LED light source, a linear polarizer, and a monochrome 1.3 MPx CCD camera. The CCD camera is interchangeable with a lens system and a spectrometer (Ocean Optics USB-2000), which has been calibrated using a low pressure krypton calibration light source with traceable spectral lines. The microscope is equipped with infinity-corrected objectives with magnifications in the range 5x to 50x. In another setup, the microscope is adapted by introducing a second beam splitter cube (50/50) just over the objective, and one of the beams is focused into a fiber connected to a spectrometer. A more detailed description of this system can be found in the Supplementary Materials. In such a configuration one can obtain the image and spectrum simultaneously, but at the cost of about 4 times reduction in intensity.

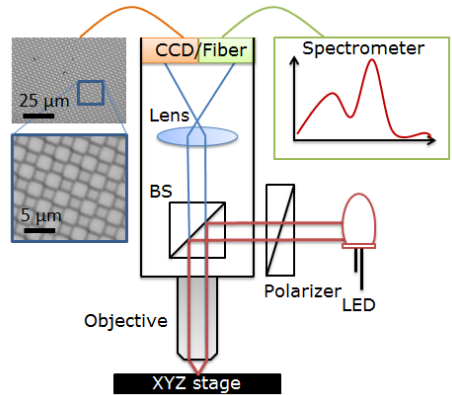

Figure 2. Sketch of experimental setup. Data acquisition can be performed using either a CCD camera or a spectrometer. The images to the left shows a calibration artefact with 3 μm pitch acquired with a 50x objective. CCD = charge-coupled device (1.3 MPx), LED = light emitting diode (5 W), BS = beam splitter cube (50/50).

The focus can be adjusted by either moving the lower part of the objective or by moving the sample stage, but in general, only the sample stage is used for focusing. The sample is brought into focus by monitoring it on the CCD detector, which is a huge advantage to other scatterometry setups, where one often struggles with finding the area of interest and bringing it in focus. Often the focus point is found on scatterometers by finding the maxima for the intensity of the spectrum, which is not a very reliable method. The CCD is also used to estimate the effective spot size for the different objectives by measuring on calibration artefacts. With the 5x objective the spot size is 1.5 mm, with the 27x objective the spot size has a diameter of 250 μm, and with the 50x objective the spot size is 125 μm.

A reference and a dark spectrum are acquired before measuring on the sample of interest. The reference spectrum, $I_{ref}(\lambda)$, is acquired on a surface with known reflection coefficients, e.g. a Si(100) substrate. The acquisition time is set to take full advantage of the dynamic range of the spectrometer. A typical acquisition time is 5 ms, which is then kept constant for this study. Secondly, a dark spectrum, $I_{dark}(\lambda)$, is acquired by removing the reference sample. The dark spectrum corrects for noise in the spectrometer and eventual ambient light conditions that give a constant signal. The diffraction efficiencies are calculated for each wavelength using

$$\eta(\lambda) = \frac{I_{sample}(\lambda) - I_{dark}(\lambda)}{I_{ref}(\lambda) - I_{dark}(\lambda)} R(\lambda), \qquad (1)$$

Where $R(\lambda)$ are the reflection coefficients of the reference sample.

To simplify the modelling only the $0^{th}$ order reflection is measured. This gives rise to the following constraint on the grating period $d$:

$$d \lesssim \frac{\lambda_{min}}{2\sin(2\theta_{NA})} \qquad (2)$$

Here, $\lambda_{min}$ is the minimum wavelength measured with the spectrometer and $\theta_{NA}$ is the collection angle for an objective with numerical aperture $NA = \sin(\theta_{NA})$ in air. In contrast to conventional imaging, it is thus favorable to use an objective with a low numerical aperture, and for the measurements presented in this study we use a 5x objective with NA=0.14. For this objective and with the cut-off wavelength $\lambda_{min}$ = 445 nm, we find that the period of the grating should be less than 803 nm to avoid $1^{st}$ order reflections from being imaged. However, as demonstrated experimentally in this paper, $1^{st}$ order reflections can to a good approximation be omitted in the simulations also for gratings with a larger pitch.

**Inverse modelling**

The data analysis is based on an inverse-modelling approach where scattering intensities are modelled first and afterwards compared to the experimental values. The scattering intensities are calculated using the rigorous coupled-wave analysis (RCWA) method as described in Refs. [22], [23]. The periodic grating is divided into slabs, for which scattering is modelled individually, and then the scattering from each slab is coupled through boundary conditions. A database including variations in height, filling factor, and sidewall angle is modelled for each pitch $\alpha=(\alpha_{height},\alpha_{FF},\alpha_{sw})$. Other methods for modelling, such as finite element analysis, can also be used, but are significantly slower. For ODM many wavelengths are needed for each model, and hence the computation time is very long using finite element methods. All models, regardless of the used method, need prior knowledge of the index of refraction and the extinction coefficients of the considered materials.

As a regularizing measure, the modelled scattering intensities are stored in a database and then compared to the experimental data using a least-squares optimization as a gauge for the quality of the fit:

$$\chi^2 = \sum_{i=1}^{N} \left[\frac{\eta - f_i(\boldsymbol{\alpha})}{\sigma_i}\right]^2 \qquad (3)$$

Here $\sigma_i$ are the uncertainties on the experimental data as described in Ref. [4], and $f_i(\boldsymbol{\alpha})$ are the modelled scattering intensities for the *i'th* element with the shape $\boldsymbol{\alpha}$. The database element with the lowest $\chi^2$ value is the best match to the model. From Eq. (3) it can be seen that experimental data points with an associated large uncertainty give a smaller contribution to the sum than data points with a relatively small uncertainty. Thus, the best model is found with most weight on the data points with the smallest uncertainty.

For increased precision, the parameters of the best-fit model are further optimized with a linear fit using neighbor diffraction efficiencies. The optimization is also used to estimate the uncertainty on the fitted parameters. It has also been demonstrated that one can use a two-step optimization procedure, where first a global and then a local optimization is applied [24].

For validation of the system a set of thin film transfer standards has been measured with the scatterometer. The transfer standards consist of $SiO_2$-coated Si(100) substrates, where the $SiO_2$ has a thickness in the range 6 nm to 1 µm. Optical constants for input to the simulations are obtained from [25]. The results are summarized in Tab. 1 and experimental data and fit can be found in Supplementary Materials. The confidence limits for the scatterometry fits are found using constant chi-square boundaries [26]. For ν degrees of freedom the chi-square distribution $\Delta\chi_\nu^2 = \chi_\nu^2 - \chi_{min}^2$ is found, where $\chi_{min}^2$ is the global minimum of the chi-square distribution. To find the confidence limit of a single parameter, ν=1, with a confidence interval of 95 %, the relation $\Delta\chi_\nu^2 < 4$ should be fulfilled. It should be stressed that the confidence limit only gives an uncertainty estimate of the parameters included in the analysis. It is from the data in Tab. 1 seen that the scatterometer measures a value within the expanded uncertainty interval of the transfer standards for thicknesses above 160 nm.

| Reference [nm] | Scatterometer [nm] |
|---|---|
| 6.0 ± 1.1 | 11 ± 8 |
| 69.7 ± 1.3 | 72 ± 4 |
| 163.2 ± 1.5 | 162 ± 4 |
| 386.4 ± 2.3 | 387 ± 2 |
| 1003.0 ± 5.1 | 1002 ± 2 |

Table 1. Thin-film measurements on transfer standards. The reference values are measured using traceable spectroscopic ellipsometry and ± denotes the expanded standard uncertainty (k=2) equivalent to a confidence interval of 95 %. For the scatterometer data ± denotes the 95 % confidence interval of the fit. For transfer standards with thickness above 160 nm the scatterometer measures the thickness within the standard uncertainty of the certified reference measurements.

**Sample with nano-textured surface**

For test of the instrument on nano-textured surfaces a multi-period 1D silicon sample with 8 gratings was fabricated. The grating patterns were defined by deep ultraviolet lithography, with grating periods ranging from 700 nm to 1400 nm in steps of 100 nm. After development of the resist, the pattern was transferred to a silicon substrate by the use of inductively coupled plasma etching, where $C_4F_8$ and $SF_6$ gasses were used.

For reference characterization of height, pitch, and sidewall angle, three different techniques had to be used. Additional experimental data can be found in Supplementary Materials. The height was characterized using a traceable NX20 atomic force microscope (AFM) from Park Instruments and analyzed using the step height module in the SPIP software package (Image Metrology) for each individual line in the image. For the area with a 800 nm pitch the height was found to h = 189.1 nm with an expanded standard uncertainty (k=2) of U(h) = 1.3 nm.

The filling factor of the grating is challenging to measure with an AFM. One has to take both the tip shape and the edge shape of the grating into account [4]. Instead, the width analysis is based on scanning electron (SEM) images. Accurate measurements with an SEM are challenged by several limitations based upon the interaction of the electron beam with the sample [27]. A detailed analysis of these factors is outside the scope of this paper, but we have omitted the demand for accurate calibration of the microscope itself by performing a relative measurement to estimate the filling factor. For the area with 800 nm pitch, the filling factor of the 1D grating was found to be FF=0.477, roughly equivalent to a width of the structures of 382 nm. The uncertainty on the SEM measurements of the filling factor has been estimated to 0.007.

Measurements of the angles of the sidewalls are unreliable using normal AFM and SEM methods. Instead the sample was measured by intermittent contact mode (tapping mode) AFM with the sample tilted 23º as shown in Fig. 2(C). Profiles were now recorded perpendicular to the grooves, and the sidewall angle was estimated from four to six grooves on the average profile of the x-gradient. The side wall angle could now be measured without having to correct for the tip shape and was found to $\vartheta=90.1$ º with an associated expanded uncertainty of $U(\vartheta)=1.5$ º for the grating with a 800 nm pitch.

**Scatterometry measurements**

All gratings on the silicon sample were measured with the scatterometer, and a blank area on the same Si(100) substrate was used for the reference measurement. Before performing the measurements the area of interest was located using the CCD camera and brought into focus by adjusting the height. Measurement results for the area with 800 nm pitch obtained with TE-polarized light are shown in Fig. 3, and data for other areas can be found in Supplementary Materials. Experimental data has been obtained for the wavelength range from 445 nm to 690 nm and smoothed using a second order Savitzky-Golay filter with a frame size of 11 points. The limited wavelength range is due to anti-reflective coating of the optical components in the Navitar microscope system.

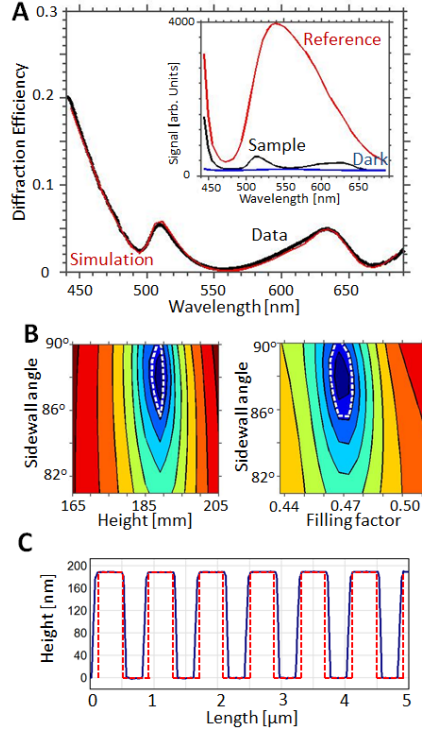

Figure 1. Scatterometry measurements for TE-polarized light on a 1D grating with a period of 800 nm etched in Si(100). (A) Experimental data (black curve) and simulation for best fit (red curve) of diffraction efficiency. The best fit is found for the parameters h=189 nm, FF = 0.468, and α=88°. The insert shows the raw data for the sample, dark, and reference spectra. (B) Color plots of the $\chi^2$ values. Dark blue shows areas with the lowest $\chi^2$ value and hence the parameters for the best fit. The dashed white curve indicates the 95 % confidence interval of the fit. (C) The profile of the best-fit data (red dashed curve) overlayered on experimental data obtained with an atomic force microscope (blue solid curve). Due to tip convolution the AFM profile overestimates width and sidewall angle.

The confidence limits for the fitting parameters can be treated individually [26], and thus the relation should be fulfilled. The $\chi^2$ values are plotted for both constant filling factor (FF=0.468) and constant height (h=189 nm) in Figure 3(B). The white dashed line indicates $\Delta\chi_\nu^2 = 4$ and hence the 95 % confidence interval of the fit. For the height, filling factor and sidewall angle the 95 % confidence interval is found to be 2 nm, 0.005, and 3°, respectively. In Figure 3(C) the profile from an AFM scan (solid blue line) has been overlayered with a profile of the best-fit data (dashed red line). It is seen that the AFM overestimates the width and sidewall angle of the structures due to the tip convolution.

Scatterometry and reference measurement results for all gratings with pitches in the range 700 nm to 1400 nm are shown in Fig. 4. The figure shows the obtained heights, fill factors and sidewall angles with $2\sigma$ confidence limits. For pitches in the range from 700 to 1200 nm there is an excellent agreement on the height and fill factor values for the different methods. For pitches above 1200 nm the agreement is worse due to the fact that the scatterometer collects a significant amount of signal from higher diffraction orders and that this contribution is not included in the scatterometry data analysis.

The sidewall angles have only been measured with AFM for the gratings with 800 nm, 1000 nm, and 1400 nm pitch, but as all structures have been dry etched in the same process, we therefore expect the sidewall angle to be close to the same value for all pitches. The scatterometry sidewall angles with pitches at 800 nm or below are within the $2\sigma$ confidence limits, whereas the sidewall angles for pitches above 800 nm are not. This clearly shows that the sidewall angle is the most sensitive parameter to the collection of signal from higher diffraction orders.

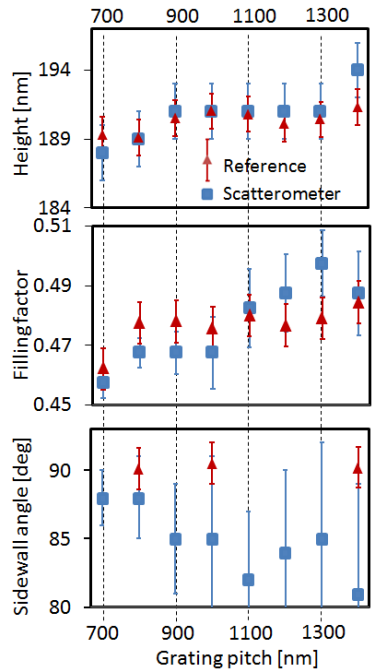

Figure 4. Height, fill factor and sidewall angle data with 2σ confidence limits for the different gratings. The scatterometry data are most accurate for pitches below 800 nm, as no higher order reflections are collected when observing these structures. The reference measurements are obtained with AFM, SEM, and tilted AFM for the height, filling factor, and sidewall angle, respectively.

**Moving the sample**

A huge advantage with the scatterometry setup is that the measurements are very robust to vibrations. This is demonstrated by moving the sample during data acquisition and summarized in Tab. 2. The sample was translated about 100 μm/s in either the x- or y-direction during data acquisition. Such a movement makes the optical image extremely blurry, but has no effect on the scattering intensities, as we still measure inside the same field on the sample. Please note that both the microscope and/or the sample can be moved during acquisition, thus making the microscope suitable to be used in a production environment.

Defocusing is detrimental for obtaining images in the image plane. However, with the scatterometer, only a very small effect on the measured values is seen, even for a defocus of 10 mm with the 5x objective. For other objectives with lower working distance the effect of defocusing is larger. Again, the sample can be moved in the z-direction during acquisition without affecting the outcome of the measurements, as long as the movement is less than 10 mm when using the 5x objective.

For the scatterometer to be integrated in an industrial production line, one will also have to take the rotation of the sample into account. For TE measurements the polarization of the incoming light should be aligned with the grating direction. We have shown that for the 800 nm grating rotations less than $6^o$ give rise to an uncertainty of less than 2 nm in the height, 0.005 for the filling factor, and $3^o$ for the sidewall angle. More details can be found in Supplementary Materials.

| Method | Height [nm] | Filling factor | Sidewall angle [$^o$] |
|---|---|---|---|
| AFM | 189.1 ± 1.3 | N/A | N/A |
| SEM | N/A | 0.477 ± 0.007 | N/A |
| Tilted AFM | N/A | N/A | 90.1 ± 1.5 |
| Scatterometer | 189 ± 2 | 0.468 ± 0.005 | 88 ± 3 |
| Scat., Moving Sample | 189 ± 2 | 0.468 ± 0.005 | 87 ± 3 |
| Scat., Defocused +10 mm | 189 ± 2 | 0.469 ± 0.005 | 88 ± 3 |
| Scat., Defocused -10 mm | 188 ± 4 | 0.470 ± 0.007 | 87 ± 3 |

Table 2. Measurements on a 1D grating with a pitch of 800 nm etched in a Si(100) substrate. Three different techniques have to be applied for reference measurements of the height, filling factor, and sidewall angle of the grating, whereas the scatterometer can measure all these in a single measurement. Moving the sample during acquisition has no effect on the scatterometry measurements, as long as one measures inside a homogeneous area. The effect of defocusing has been tested with the 5x objective.

**Conclusion**

In conclusion, we have demonstrated that by simple adaptions to an optical microscope we can measure nano-textured surfaces with a resolution in the nanometer range. The microscope has been validated by measuring on certified transfer artefact and 1D gratings with pitches in the range from 700 nm to 1400 nm. The measurements are very robust, such that vibrations of the sample and/or the microscope do not affect the results. The sample can be translated during acquisition, as long as the beam spot is kept inside an area with homogenous structures, which makes the proposed microscope well suited for implementation in a production environment.

**Acknowledgement**

The authors acknowledge the support from the Danish National Advanced Technology Foundation through the project NanoPlast, The Danish Council for Strategic Research through the project Polynano, and the European Union: Theme NMP.2012.1.4-3 Grant no. 309672.

REFERENCES

[1]     R. Leach, Fundamental Principles of Engineering Nanometrology, Elsevier, 2014.

[2]     X. Niu, N. Jakatdar, J. Bao and C. Spanos, "Specular spectroscopic scatterometry," *IEEE Transactions on Semiconductor Manufacturing,* vol. 14, no. 2, pp. 97-111, 2001.

[3]     N. Agersnap, P.-E. Hansen, J. C. Petersen, J. Garnaes, N. Destouches and O. Parriaux, "Critical dimension metrology using optical diffraction microscopy," *Proc. SPIE,* vol. 5985, pp. 68-78, 2005.

[4]     J. Garnaes, P.-E. Hansen, N. Agersnap, J. Holm, F. Borsetto and A. Kuhle, "Profiles of a high-aspect-ratio grating determined by spectroscopic scatterometry and atomic-force microscopy," *Applied Optics,* vol. 45, no. 14, pp. 3201-3212, 2006.

[5]     D.-M. Shyu, Y.-S. Ku and N. Smith, "Angular scatterometry for line-width roughness measurement," *Proc. SPIE 6518, Metrology, Inspection, and Process Control for Microlithography XXI,* vol. 65184G, 2007.

[6]     A. Kato and F. Scholze, "Effect of line roughness on the diffraction intensities in angular resolved scatterometry," *Applied Optics,* vol. 49, no. 31, pp. 6102-6110, 2010.

[7]     O. Paul, F. Widulle, B. H. Kleemann and A. Heinrich, "Nanometrology of periodic nanopillar arrays by means of light scattering," *Proc. of SPIE,* vol. 8788, p. 87881O, 2013.

[8]     P. Boher, J. Petit, T. Leroux, J. Foucher, Y. Desieres, J. Hazart and P. Chaton, "Optical

Fourier transform scatterometry for LER and LWR metrology," in *Proc. SPIE 5752, Metrology, Inspection, and Process Control for Microlithography XIX*, 2005.

[9]   J. P. Ogilvie, E. Beaurepaire, A. Alexandrou and M. Joffre, "Fourier-transform coherent anti-Stokes Raman scattering microscopy," *Opt. Lett.,* vol. 31, pp. 480-482, 2006.

[10]  S. Roy, N. Kumar, S. Pereira and H. Urbach, "Interferometric coherent Fourier scatterometry: a method for obtaining high sensitivity in the optical inverse-grating problem," *Journal of Optics,* vol. 15, no. 7, p. 075707, 2013.

[11]  N. Kumar, O. E. Gawhary, S. Roy, V. Kutchoukov, S. Pereira, W. Coene and H. Urbach, "Coherent Fourier Scatterometry (Tool for improved sensitivity in semiconductor metrology)," *SPIE Advanced Lithography,* pp. 83240Q-83240Q-8, 2012.

[12]  V. F. Paz, S. Peterhänsel, K. Frenner and W. Osten, "Solving the inverse grating problem by white light interference Fourier scatterometry," *Light Sci Appl,* vol. 1, p. e36, 2012.

[13]  S. Peterhänsel, H. Laamanen, M. Kuittinen, J. Turunen, C. Pruss, W. Osten and J. Tervo, "Solving the inverse grating problem with the naked eye," *OPTICS LETTER,* vol. 39, no. 12, p. 3547, 2014.

[14]  S. Kinoshita, S. Yoshioka and J. Miyazaki, "Physics of structural colors," *Rep. Prog. Phys.,* vol. 71, no. 7, p. 076401, 2008.

[15]  J. S. Clausen, E. Højlund-Nielsen, A. B. Christiansen, S. Yazdi, M. Grajower, H. Taha, U. Levy, A. Kristensen and N. A. Mortensen, "Plasmonic Metasurfaces for Coloration of

Plastic Consumer Products," *Nano Letters,* vol. 14, no. 8, pp. 4499-4504, 2014.

[16] B. Tian, X. Zheng, T. J. Kempa, Y. Fang, N. Yu, G. Yu, J. Huang and C. M. Lieber, "Coaxial silicon nanowires as solar cells and nanoelectronic power sources," *Nature,* vol. 449, pp. 885-889, 2007.

[17] J. Wallentin and e. al, "InP Nanowire Array Solar Cells Achieving 13.8% Efficiency by Exceeding the Ray Optics Limit," *Science,* vol. 339, p. 1057, 2013.

[18] N. Anttu, S. Lehmann, K. Storm, K. A. Dick, L. Samuelson, P. M. Wu and M.-E. Pistol, "Crystal Phase-Dependent Nanophotonic Resonances in InAs Nanowire Arrays," *Nano Lett.,* 2014.

[19] G. Grzela, R. Paniagua-Domínguez, T. Barten, D. van Dam, J. A. Saìnchez-Gil and J. G. Rivas, "Nanowire Antenna Absorption Probed with Time-Reversed Fourier Microscopy," *Nano Lett.,* vol. 14, pp. 3227-3234, 2014.

[20] P. Lalanne and E. Silberstein, "Fourier-modal methods applied to waveguide computational problems," *OPTICS LETTERS,* vol. 25, no. 15, p. 1092, 2000.

[21] M. Karamehmedović, P.-E. Hansen, K. Dirscherl, E. Karamehmedović and T. Wriedt, "Profile estimation for Pt submicron wire on rough Si substrate from experimental data," *Optics Express,* vol. 20, no. 19, pp. 21678-21686, 2012.

[22] A. M. G. Moharam, A. Pommet, E. B. Grann and T. K. Gaylord, "Stable implementation of the rigorous coupled-wave analysis for surface relief gratings: enhanced transmittance

matrix approach," *J. Opt. Soc. Am.,* vol. A12, p. 1077–1086, 1995.

[23]  B. L. Li, "Fourier modal methods for crossed anisotropic gratings with arbitrary permittivity and permeability tensors," *J. Opt.A Pure Appl. Opt.,* vol. 5, p. 345–355, 2003.

[24]  P.-E. Hansen and L. Nielsen, "Combined optimization and hybrid scalar–vector diffraction method for grating topography parameters determination," *Materials Science and Engineering: B,* pp. 165-168, 2009.

[25]  E. D. Palik, Handbook of Optical Constants of Solids, Boston: Academic Press, 1985.

[26]  W. H. Press, S. A. Teukolsky, W. T. Vetterling and B. P. Flannery, Numerical Recipes in C++, Cambridge: Cambridge, 2002.

[27]  M. T. Postek, "Critical Issues in Scanning Electron Microscope Metrology," *J.Res. Natl. Inst. Stand. Technol.,* vol. 99, no. 5, p. 641, 1994.